\documentclass[manuscript=article]{achemso}
\usepackage[version=3]{mhchem} 
\usepackage{amsmath}
\usepackage{natbib}
\usepackage{subcaption}
\usepackage{xr}
\usepackage{xcolor}
\usepackage{ulem}
\usepackage[colorinlistoftodos, textwidth=20mm]{todonotes}
\author{Andreas A. Hennig}
\affiliation{PoreLab, Department of Physics, Norwegian University of Science and Technology, Høgskoleringen, Trondheim, N-7491, Norway.}
\author{Ilaria Beechey-Newman}
\affiliation{PoreLab, Department of Physics, Norwegian University of Science and Technology, Høgskoleringen, Trondheim, N-7491, Norway.}
\author{Natalia Kizilova}
\affiliation{Warsaw University of Technology, Institute of Aeronautics and Applied Mechanics, Nowowiejska, 24, Warsaw, 00-665, Poland.}
\alsoaffiliation{V.N. Karazin Kharkiv National University, Department of Applied Mathematics, Svobody sq., 4, Kharkiv, 61-022, Ukraine.}
\author{Erika Eiser}
\affiliation{PoreLab, Department of Physics, Norwegian University of Science and Technology, Høgskoleringen, Trondheim, N-7491, Norway.}
\email{erika.eiser@ntnu.no}

\title{Arc-length characterization of finite, radial growth patterns}

\begin{document}


\begin{abstract}
We  present a method to characterize the distribution of length-scales of finite, disordered patterns with, on average, radial symmetry.
This method makes it possible to quantify the distribution of characteristic length scales in cases where the conventional ``linear'' chord method does not work.
We show that the method can clearly distinguish regular patterns, patterns that are formed by diffusion-limited aggregation, and patterns that form during the slow drying of confined, colloid-laden droplets, explained by Beechey-Newman {\it et al.}~\cite{Beechey-Newman_etal_2025}.
We also introduce a method to find the centre-point of these finite patterns, without assuming a full connectivity in the pattern.
The method should be widely applicable to other, finite quasi-two-dimensional patterns like dendritic structures, viscous fingering, liquid crystal patterns and bacterial growth.
\end{abstract}


\section{Introduction}
\begin{figure}[!ht]
\centering
\includegraphics[width=0.45\linewidth]{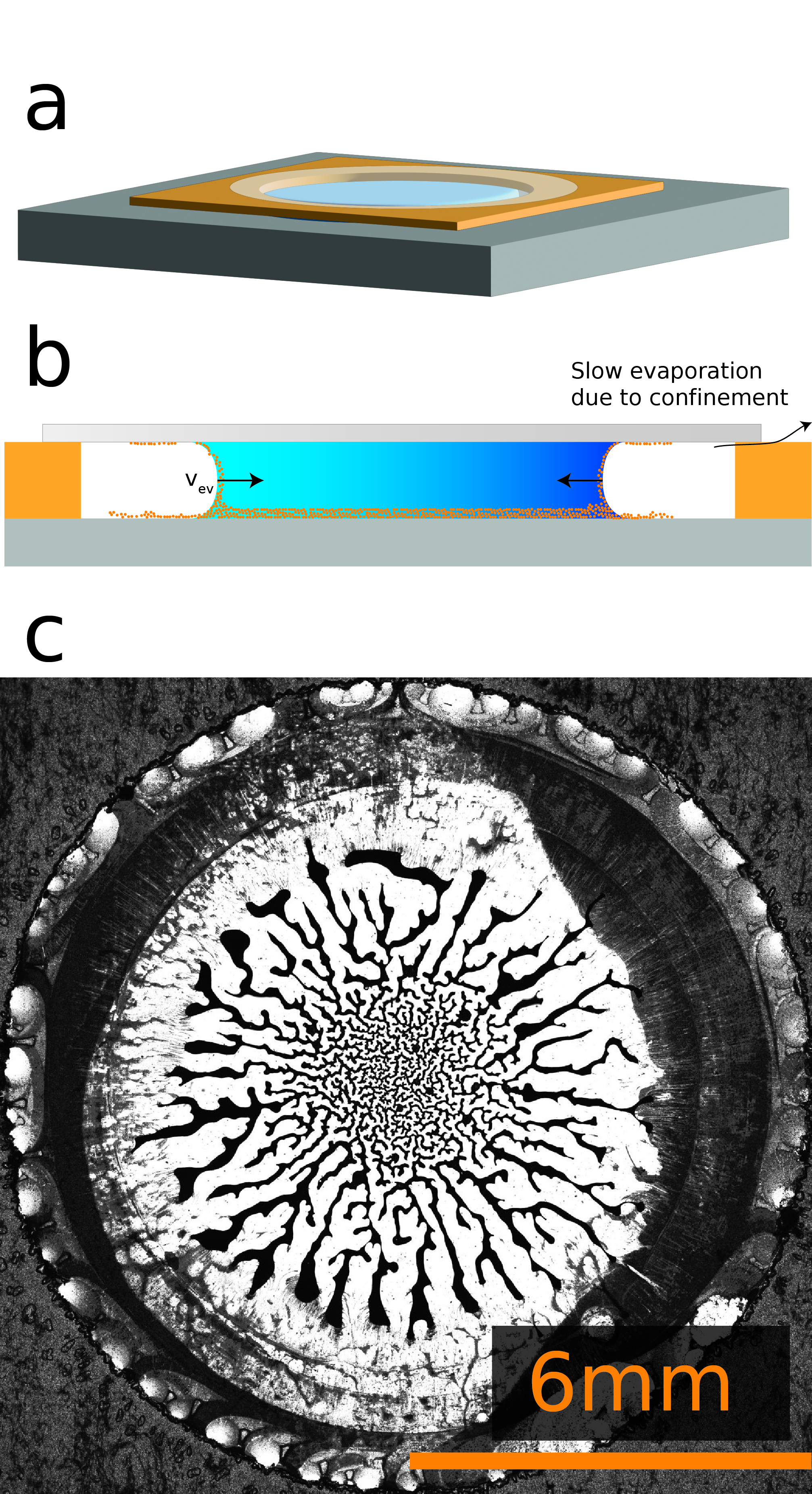}
\caption{Illustration of the experimental setup from ref.\,\citenum{Beechey-Newman_etal_2025} that was used to obtain the colloidal maze-like patterns. A microscope slide is prepared with a double-sided sticky tape, where we punch a hole of $r=6$ mm and deposit a droplet of water containing 1\,wt.\% of 1 $\mu$m large TPM-particles, as shown in (a). Then, a cover-slip is placed on top of the circular cell, forcing the evaporation process to take many days. A cross-section of the cell is shown in (b), illustrating the colloid deposition both on the bottom and top glass plates. The slow evaporation process drives colloids to assemble into a contiguous maze, as seen in the photograph shown in (c).} \label{fig:Experimental Setup}
\end{figure}

The so-called ``chord-length distribution method''~\cite{Mering_Tchoubar_1968} is a convenient tool to characterize the distribution of length scales of clusters and voids in a disordered yet homogeneous, two-dimensional pattern of two phases ({\it e.g.} solid and air). 
This chord analysis aims to determine the distributions of segment-lengths of a (any) straight line that intersects a large number of solids and void spaces, {\it e.g.} in a gel network~\cite{Zupkauskas_etal_2017}.
The conventional chord analysis does not provide meaningful results for patterns that are not homogeneous and finite.
Yet such patterns are important: many quasi-two-dimensional aggregation phenomena yield patterns that are i) finite and ii) not homogeneous in space.
One example among many is the pattern that is formed by diffusion limited aggregation (DLA).
For such, on average radial patterns, chord analysis does not provide a meaningful description.
Here we propose a closely related approach, to probe the arc-length distribution of circles that intersect both solid and void domains.
Importantly, this arc-length distribution depends on the radius of the intersecting circle, even when averaged over many similar patterns.
In what follows, we will refer to this type of analysis as circular chord-length analysis: this name is less precise than arc-length distribution, but stresses the analogy with the conventional chord-length distribution.
Arc-length distributions can serve as a ``fingerprint'' of disordered radial patterns, obtained experimentally as shown in Figure\,\ref{fig:Experimental Setup}.
We compare the circular arc-length distribution for three visually different patterns (3 typical colloidal drying patterns; a diffusion limited aggregation (DLA) pattern; and a ``random birthday pie'' pattern) to illustrate that the method can be used to quantify the visual differences between these patterns.
\section{Systems}
Roughly radial patterns often form due to the deposition of colloids, either due to evaporation of the solvent, or due to (diffusion-limited or reaction-limited) aggregation of colloids from a colloid-laden suspension. 
A classic example is the famous ``coffee-stain-effect''~\cite{Deegan_etal_1997a} where the  dispersed coffee grounds mostly accumulate in the rim of the dried stain.
The effect was explained in ref.~\citenum{Deegan_etal_1997a} for the case of droplets on partially wetting surfaces, where the droplet circumference gets pinned.
As the droplet evaporates, there will be a capillary flow of the solvent towards the rim, transporting almost all colloidal particles towards the perimeter. In inkjet printing, but also in coatings and paintings, this effect is unwanted, and  efforts have been made to overcome the capillary flows that drive deposition on the pinned contact line~\cite{Hu_Larson_2006a,Abdulsahib2022,Majumder_etal_2012,li_et_al_2020}. 

More recently, maze-like patterns of slowly evaporating cylindrical droplets were described by Beechey-Newman {\it et al.}~\cite{Beechey-Newman_etal_2025} 
These patterns form inversely-ramified patterns, meaning that the branches in the pattern become thinner towards the centre.
In the present paper, we aim to determine and compare  ``fingerprints'' of the various patterns listed above, with those of simpler radial patterns. 

The fractal dimension of a pattern is a convenient quantity to characterize  naturally-formed radial patterns that are scale-invariant. 
This fractal dimension can be used to distinguish different growth mechanisms; {\it e.g.}, viscous fingering, stable displacement or capillary fingers in the invasion percolation of two immiscible fluids in porous media \cite{Hansen_etal_2022}. 
In contrast, the patterns observed in ref.~\citenum{Beechey-Newman_etal_2025} are {\it not}\; scale-invariant: the characteristic length scale of the branches changes rather abruptly with radius, something that does not lend itself to a fractal or multi-fractal description. 
It is for this reason, that for such systems, we explore the angular equivalent of the chord method.

The remainder of this paper is organized as follows: first we introduce the circular arc-length distribution method (CAL), and its relation to the linear chord-length distribution introduced by M\'{e}ring and Tchoubar~\cite{Mering_Tchoubar_1968}. 
We then consider examples of applications of the CAL method to experimentally observed drying patterns of colloidal droplets, for DLA patterns,  and for and artificial radial patterns.
We show that, for experimental patterns,  the CAL method can generate unique fingerprints that depend on the radius of the intersecting circle. 
Such an analysis is particularly useful in cases where the nature of the pattern formation changes with the radius of the deposit.

\section{Arc-length distributions}

\begin{figure}
\centering
\includegraphics[width=0.75\linewidth]{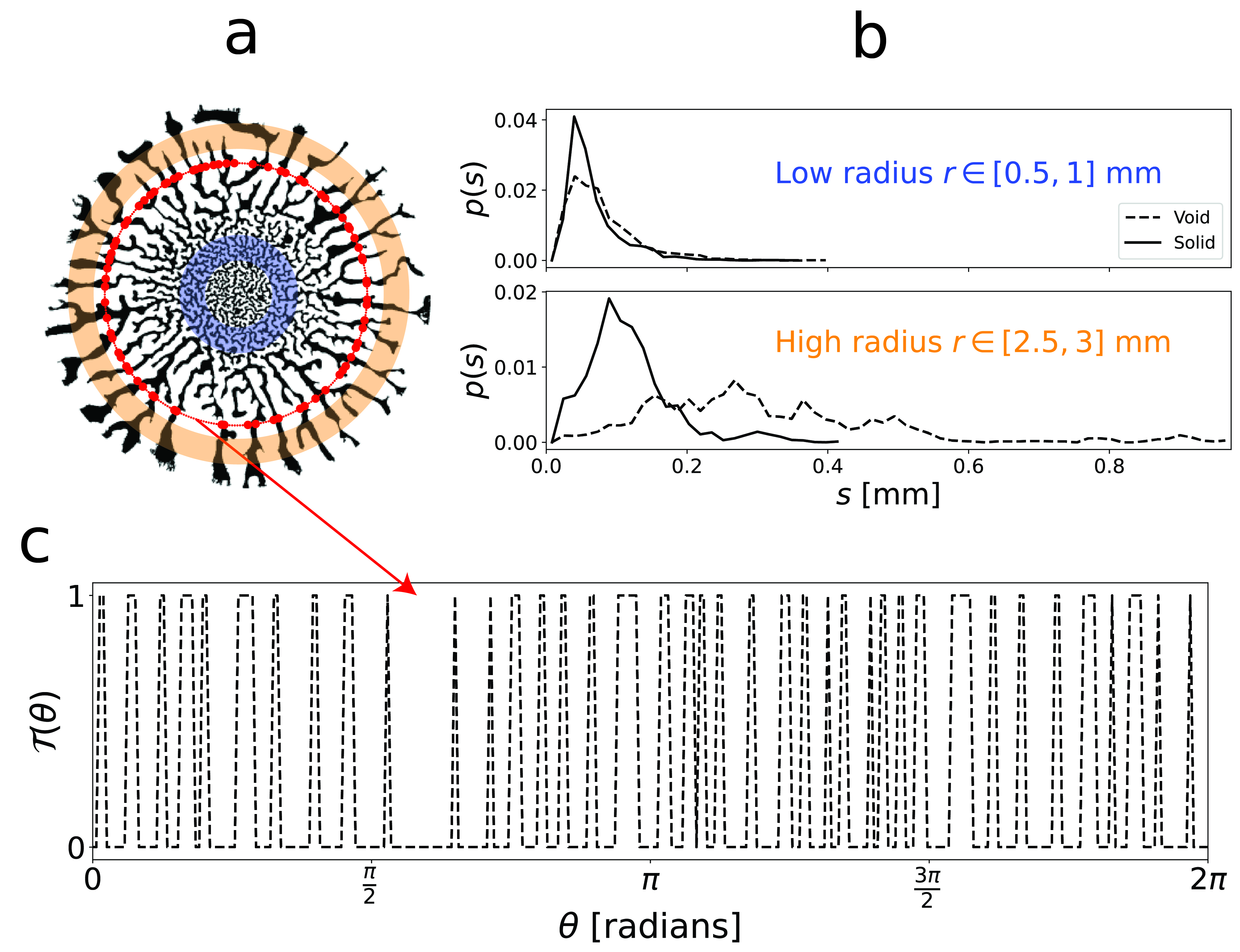}
\caption{(a) Illustration of the circular arc-length distribution for the colloidal drying pattern P2, that was appropriately cropped and binarized. For a given radius (red circle), each void-solid (white-black) interface is marked with a red dot, allowing us to store the length along the circle between each interface. (b) Resulting distribution of the arc lengths $s$ between interfaces, where $p(s)$ is the probability for a given $s$. Radii lying in the orange shell show a broader distribution than those lying in the blue, inner shell, with finer structures. (c) Telegraph signal $\mathcal{T}(\theta)$ obtained from the particular radius illustrated in (a), with $\theta$ starting from the 3 o' clock position moving clockwise. In the signal, 1 indicates the solid phase and 0 corresponds to the void phase.
The details of the CAL distribution differ from sample-to-sample, but the overall characteristics remain the same.}
\label{fig:Fig_3}
\end{figure}

The circular arc-length analysis used in this work is a fairly straightforward generalization of the linear chord-length method of Levitz and Tchoubar~\cite{Levitz_Tchoubar_1992}.
In its simplest form, the CAL method is used to characterize biphasic patterns, where the two ``phases'' might be real phases of matter, {\it e.g.}, solid and vapor, or simply different labels ({\it e.g.}, black and  white, characterizing an artificially generated pattern), as seen in Figure\,\ref{fig:Fig_3}. 

We use a fairly standard method to binarize the experimental images, similar to the method used in Zupkauskas {\it et al.}~\cite{Zupkauskas_etal_2017}.
First, we apply a Gaussian blur on the image, and the threshold it using Otsu's method~\cite{Otsu_1979}.
Finally, a labelling algorithm is used to filter out all clusters below a certain size, to minimize the random labelling errors that inevitably occur.

The first step in the CAL analysis is to determine the centre of the pattern.
For natural patterns such as DLA, the choice of the centre is obvious: it is the point where the pattern originated. 
Conversely, in the case of drying patterns, the centre might be identified with the last point to dry; ideally circles around the central point should correspond to the location of the drying front at different points in time. 
If the drying proceeds symmetrically, then the last point to dry would  be the correct choice for the central point in the CAL analysis, but in general, there is some arbitrariness in choosing the centre of the pattern.

Once the centre of the pattern has been located or chosen, we can draw circles of increasing radius $r$ around this centre.
Typically, such a circle will intersect many boundaries between the two phases (say: ``solid'' and ``void'). 
If we attribute the value $1$ to all points in the solid phase, and $0$ to points in the void phase, then we can construct a function $\mathcal{T}(\theta, r)$, where $0<\theta\le 2\pi$ is the angle specifying a point on the circle corresponding to radius $r$. 
As $\mathcal{T}(\theta, r)$ can have only the two values $0$ and $1$, the resulting function at constant $r$ looks like a (random) telegraph signal that depends on $\theta$. 
In experiments, the resolution with which the pattern is obtained is limited by the pixel size, typically a square with length and width $\Delta$. 
A pixel ``belongs'' to a given circle if that circle intersects the pixel area.
In general, a pixel might belong to more than one circle. 
However, if the radii of adjacent circles differ by more than $\Delta\sqrt{2}$ then every pixel belongs to at most one circle.
Of course, the density of pixels on a circle will, in general, not be constant, but in practice that  is not a serious problem for determining arc-segment lengths.

In some cases, pixels may have the wrong ``label'' due to noise. 
Unless the noise level is very high, it is  unlikely that two or more adjacent pixels will be incorrectly labeled due to noise.
To remove incorrectly labeled pixels, one should check if its label differs from that of all its neighbors, {\it i.e.}, if the pixel is isolated.
Such pixels should be removed from the analysis. 
Of course, this procedure can be refined if necessary. 
However, a labeling error at a domain boundary cannot be removed in this way.
In the cases that we study, the noise levels are so low that such errors should have a negligible effect.

Having thus identified the ``random telegraph'' signal on a given circle, we can identify arc-segments on the circle where the label remains the same.  
The arc-length of a segment is  defined as the distance in the angle $\theta$ between two adjacent label changes. 
Note that, as the pattern in $\theta$ is periodic with a period $2\pi$, the distance between the two boundaries of an arc segment should be defined as the distance (in $\theta$) between a given segment boundary and the {\it nearest periodic image} of the adjacent boundary.

\section{Analysis of experimental and synthetic patterns}

\begin{figure*}[ht]
\centering
\includegraphics[width=0.75\linewidth]{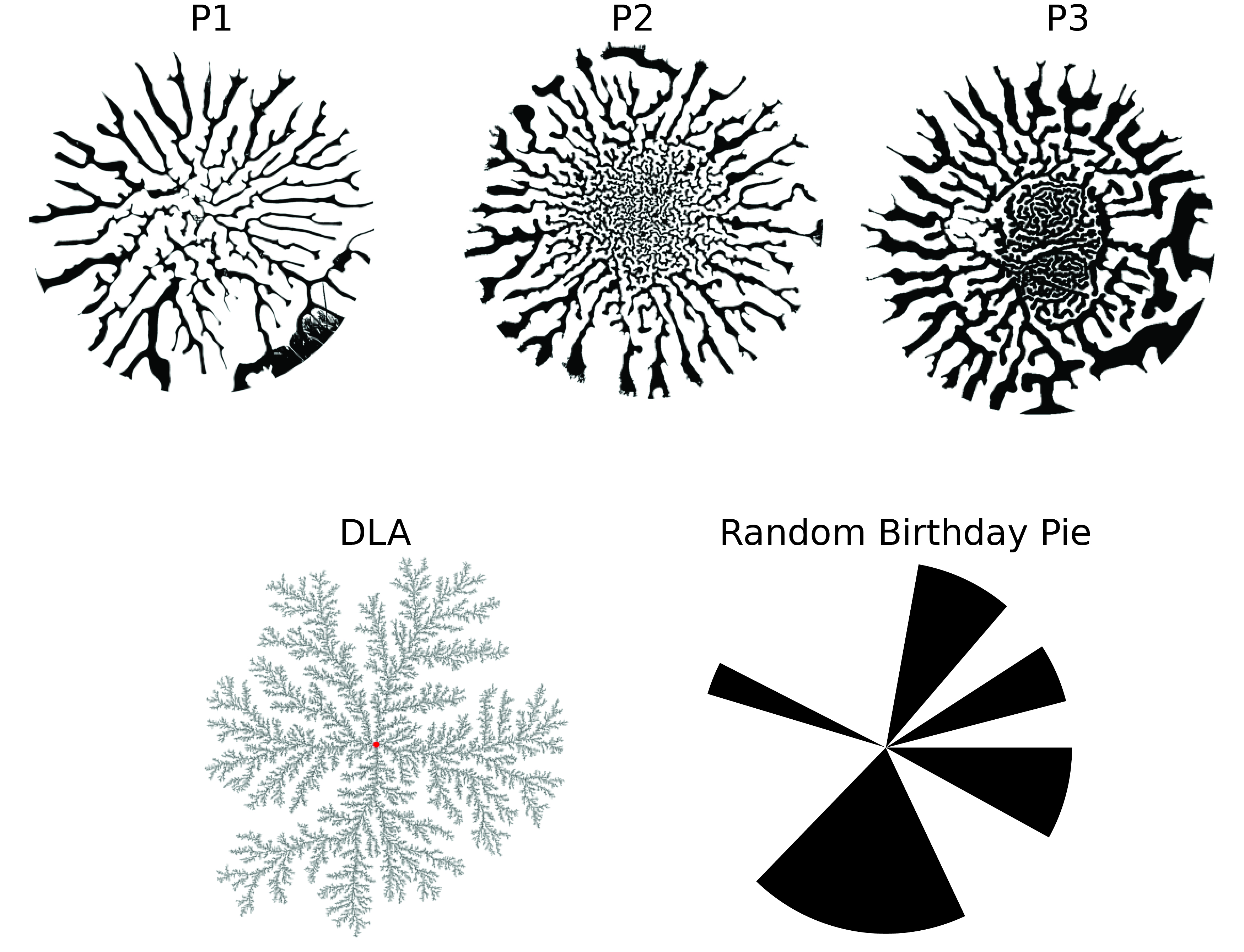}
\caption{The 5 patterns we compare in this paper, where 3 are drying patterns, obtained by the method described in Figure\,1\,\cite{Beechey-Newman_etal_2025}. The images were then appropriately cropped and binarized. The lower two images are that of a DLA pattern, which was generated with 400000 particles in a $2048 \times 2048$ grid and a RBP pattern, which produces a radially invariant CAL distribution. 
} \label{fig:Pure Patterns}
\end{figure*}
We tested the CAL method by applying it to the quasi-2D colloidal drying patterns P1, P2 and P3 in Figure\,\ref{fig:Pure Patterns}.
Although these patterns were formed under similar conditions, they can differ qualitatively as explained by Beechy-Newman et al.\cite{Beechey-Newman_etal_2025}.
In addition, we applied the same CAL analysis to a numerically generated, two-dimensional diffusion-limited aggregation model~\cite{witten_diffusion-limited_1981} and an artificial pattern of random segments, which we refer to as the ``random birthday pie'' (RBP).
As was discussed by Levitz and Tchoubar~\cite{Levitz_Tchoubar_1992}, 
DLA patterns yield a power-law distribution of the arc lengths in the void phase: $p_p(s)\sim 1/s^{d_m-1}$ where $s$ is the arc length and $d_m$ is the fractal dimension of the pattern in the void phase. 
Below, we first apply the circular arc-length analysis to a radial colloidal drying pattern~\cite{Beechey-Newman_etal_2025} and then compare it to that of a DLA pattern.
As the mechanisms by which the drying and DLA patterns form are very different, we might expect that this should result in rather different CAL distributions.

The circular arc-length distributions of the drying pattern P2 are presented in Figure\,\ref{fig:Fig_3}. 
From this figure we see that the distributions will vary with the radius of the pattern; the goal of the method is to understand these variations. It is instructive to compare our patterns with two characteristic behaviours. In DLA, the finger thicknesses will be 1 pixel, with only a few exceptions. Conversely, a radially invariant pattern, such as the RBP, will have a finger thickness proportional to the radius with a constant determined by the fraction of solid phase in the pattern: $s = \varphi \cdot r$.
 
\begin{figure}
\centering
\includegraphics[width=0.75\linewidth]{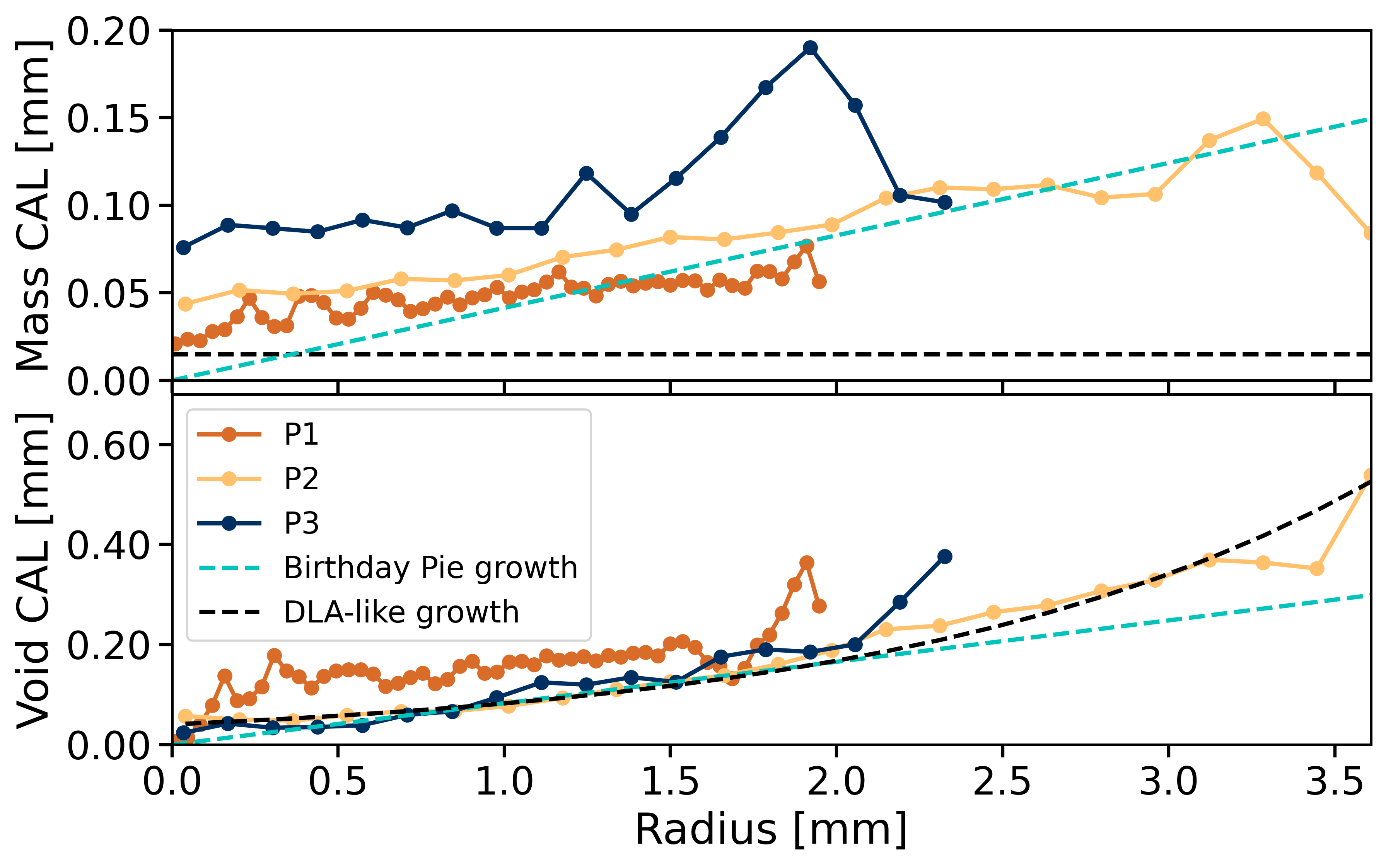}
\caption{Average arc-lengths of both phases, averaged over $\mathrm{d}r=20$ pixels, for the 3 drying patterns P1-P3. The growth trends of the synthetic patterns are added as dashed lines for illustration.}\label{fig:Mean CAL}
\end{figure}

In the conventional chord length analysis, where we create histograms of cord-lengths in $x$- and $y$-directions in a porous medium, for instance, we are usually interested in finding an average characteristic length scale. This could be for instance the average pore-size in the system \cite{Zupkauskas_etal_2017}. 
In our radial analogy, we illustrate the mean arc lengths for the experimentally obtained and artificial patterns in Figure\,\ref{fig:Mean CAL}. The RBP pattern scales linearly in both phases (with the solid fraction $\varphi\approx 0.5$ as the slope; turquoise dashed lines). In contrast, the DLA pattern has a constant solid arc length, while the average {\it void} arc-length grows as a power-law with the radius (black dashed lines).
We find that the drying patterns fall somewhere between the two synthetic patterns. 
All colloidal drying patterns have on average thicker fingers ({\it solid phase}) and thicker voids ({\it void phase}) at the outer radii. 
The main difference is that the finger thickness and void thickness are comparable close to the centre, but the average void finger thickness grow much faster. 

\begin{figure}
\centering
\includegraphics[width=0.75\linewidth]{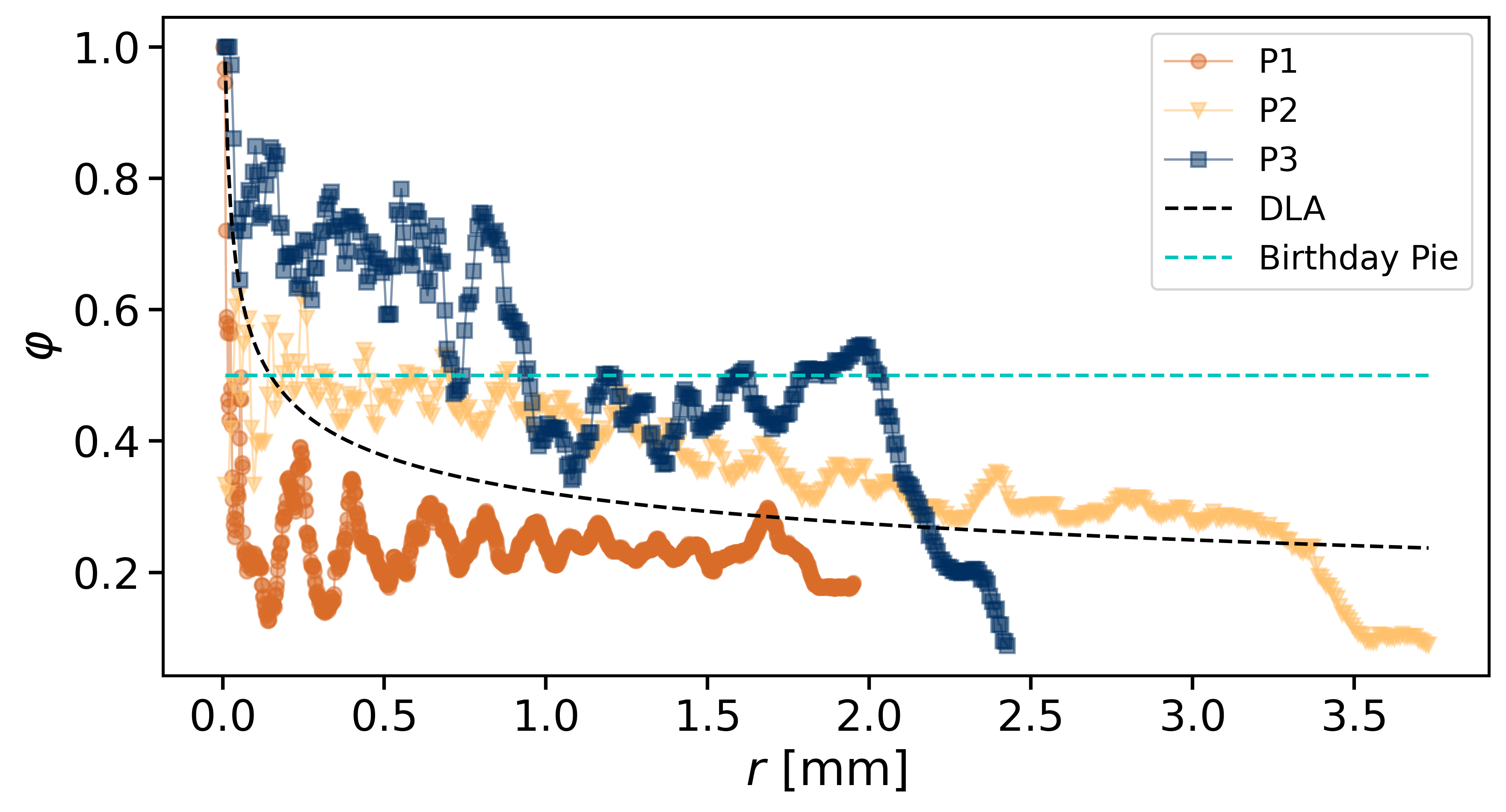}
\caption{Solid fraction $\varphi$ along the circular shell of $dx=1$ at radius $r$. 1 means the shell contains only solid, {\it e.g.} at the centre of the patterns, while 0 meaning the shell contains only void phase.}\label{fig:area fractions}
\end{figure}

We can also explore other properties of the telegraph signal we find in our radial chord length analysis.
Starting with the smallest binning length of $\mathrm{d}r=1$ pixel, we study three properties with respect to the radius: Area fractions, number of interfaces and cross-radius correlation functions. 
In Figure\,\ref{fig:area fractions}, we present the fraction of the solid phase ({\it i.e.} by counting the frequency of 1 in the telegraph signal) of the drying patterns at each radius.
We expect $\varphi=0$ far away where we have a monolayer deposition and therefore do not have any clear fingers \cite{Beechey-Newman_etal_2025}; likewise we expect that the centre of the pattern should be in the solid phase, and expect $\varphi=1$ as $r\rightarrow 0$.
All 3 colloidal maze patterns studied display 3 different characteristics in $\varphi(r)$: P1 oscillates around a stable value, where the peaks in the oscillations correspond to the radii at which branching occurs; in pattern P2 the solid fraction increases as the pattern dries; pattern P3 has two fairly sharp jumps with a stable value between them.
As these patterns are formed in similar conditions, the fluid instability offer a surprising sensitivity to the subtle differences in the initial conditions.

\begin{figure}
\centering
\includegraphics[width=0.75\linewidth]{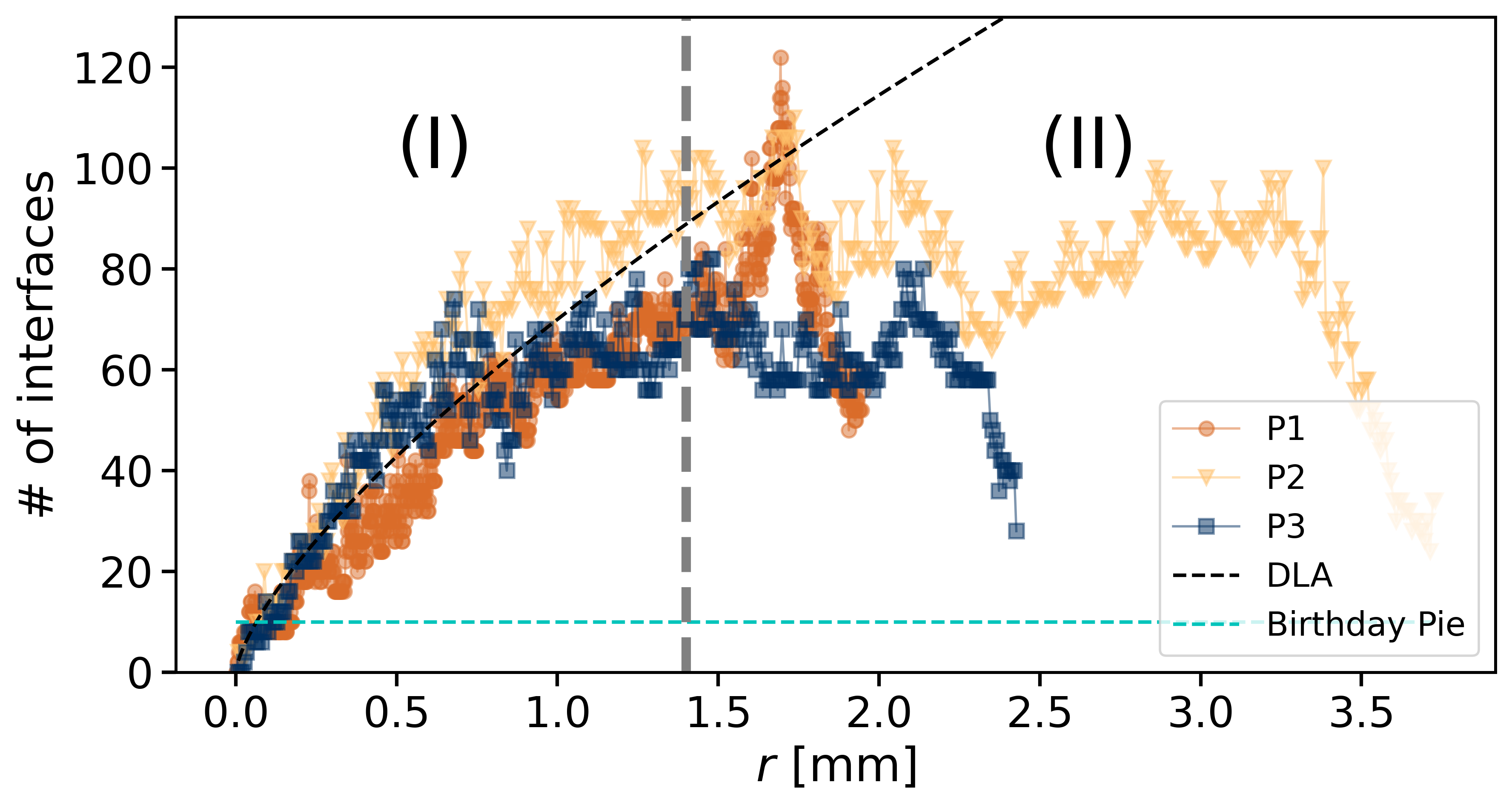}
\caption{Number of phase shifts of the telegraph signal, corresponding to the surface area the fingers generate. In a DLA, the number of interfaces grows with the circumference, while radially invariant patterns, {\it e.g.} random birthday pies, have a constant number of phase shifts. We mark this figure with two regions where the colloidal patterns exhibit different scaling: Region (II) is characterised with a fairly stable number of interfaces; at some point, the fingers start merging faster than they are formed, and in region (I) there is a fairly stable merging rate across the drying patterns.}\label{fig:surface areas}
\end{figure}

At each radius, it is possible to count the number of discontinuous jumps in the telegraph signal.
Each of these jumps represents a change in the phase ({\it i.e.} an interface); the lengths between phase changes are the arc lengths we study.
Thus, counting the number of jumps in the signal is the same as counting the number of distinct arc lengths, and is therefore proportional to the number of fingers, at each radius.
The scaling is known for the DLA and Birthday Pie: DLA creates interfaces as a power-law with its fractal dimension $N\sim r^{(D_f-1)}$,~\cite{kaufman_parallel_1995} while a radially invariant pattern has the same number of interfaces at every radius. 
The phase jumps of our drying patterns are shown in Figure\,\ref{fig:surface areas} and display a transition between the characteristic behavior of DLA, close to the centre, and RBP far away from the centre.
The fingers clearly merge as the droplet dries, while the fluid instability causing the fingers also form new fingers during drying.
The scaling in region (II) in Figure\,\ref{fig:surface areas} means the rate of merging roughly matches the rate of forming new fingers.
At some point there is a transition where the fingers merge more frequently, leading to region (I), where the colloidal fingers follow a linear "DLA-like" decay towards the centre.

\begin{figure}
    \centering
    \includegraphics[width=0.75\linewidth]{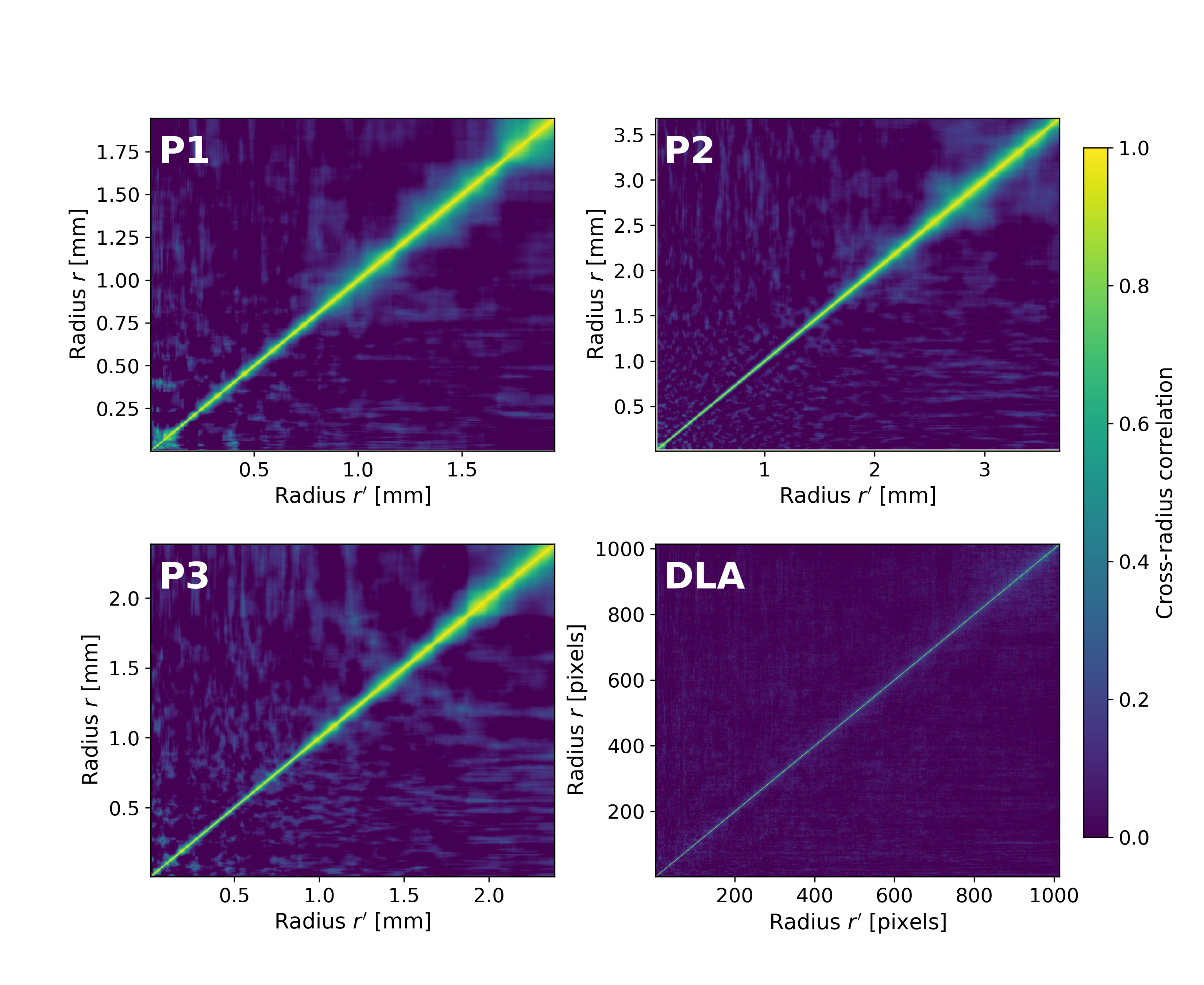}
    \caption{Telegraph signal correlation matrix for patterns P1-P3 and DLA.}
    \label{fig:Correlation_Matrix}
\end{figure}

\begin{figure}
    \centering
    \includegraphics[width=0.75\linewidth]{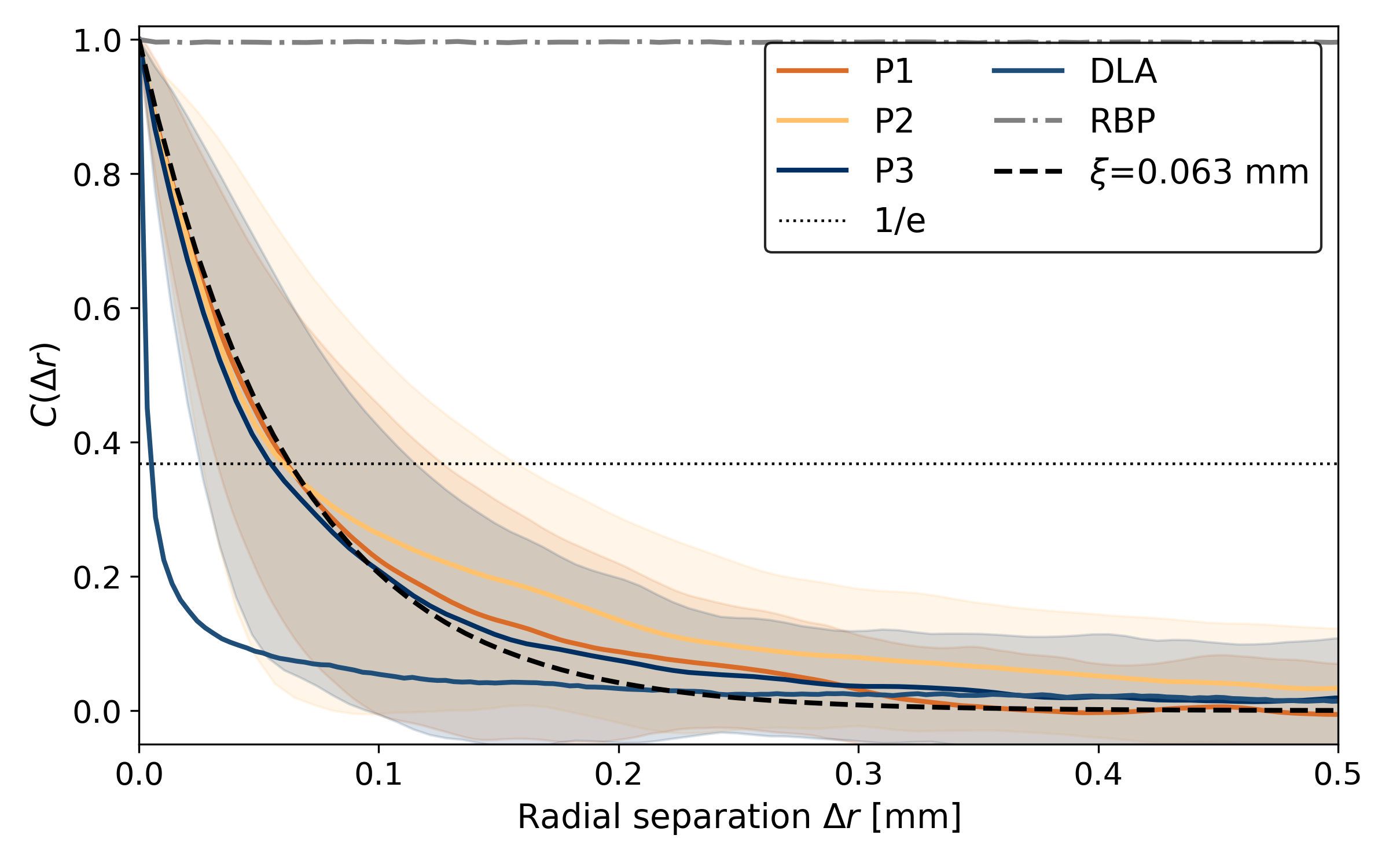}
    \caption{Average correlations of telegraph signals, by separation distances $\Delta r$ for all the 5 patterns. The RBP remains perfectly correlated throughout the pattern, while the DLA decorrelates rapidly -- with a correlation length of 2 pixels. The drying patterns P1-P3 decay exponentially, with decay lengths of 0.063, 0.065 and 0.061\,mm. We plot the average decay length of $\xi=0.063$\,mm.}
    \label{fig:Average_correlation_separation}
\end{figure}

We may also study the correlation between telegraph signals, to quantify the structural similarity between radii.
Here, we want to factor out the correlation between solid fractions $\varphi$; we therefore remove the average from the telegraph signal and only look at the fluctuating part $\widetilde{\mathcal{T}}(r, \theta) = \mathcal{T}(r, \theta) - \langle \mathcal{T}(r, \theta)\rangle_\theta$, with $\langle A(\theta \rangle_\theta=\frac{1}{2\pi}\int_0^{2\pi} A(\theta) \mathrm{d}\theta$.
The cross-correlation function of the pattern is defined as
\begin{equation}\label{eq:Correlation}
    C(r, r')=\frac{\left\langle \widetilde{ \mathcal{T}}(r,\theta)\widetilde{ \mathcal{T}}(r',\theta)\right\rangle_\theta}
    {\sqrt{\left\langle \widetilde{ \mathcal{T}}(r,\theta)^2\right\rangle_\theta
    \left\langle \widetilde{ \mathcal{T}}(r',\theta)^2\right\rangle_\theta}}.    
\end{equation}
A pattern that is radially invariant, such as the RBP patterns, will have the property that $\mathcal{T}(r, \theta)=\mathcal{T}(r', \theta)$; these are perfectly correlated at all radii, with only a small numerical error due to rasterization of the image.
The correlation for the DLA and P1-P3 patterns are shown in Figure\,\ref{fig:Correlation_Matrix}. 
We can see that the drying patterns P1-P3 decorrelate faster close to the centre, and are more persistent in the outer regions, where the fingers initially form. 
We then use equation \eqref{eq:Correlation} to find the average correlation of the separation of the compared signals as $C(\Delta r)=\big\langle C(r,r')\big\rangle_{|r-r'|=\Delta r}$.
We plot these in Figure\,\ref{fig:Average_correlation_separation}, where the decorrelation of the 3 drying patterns collapse to the same exponential decay curve. We introduce the persistence length $\xi$ as the length where the signal has decorrelated by a factor $1/e$, {\it i.e.} $C(\Delta r)=\exp(-r/\xi)$.
The persistence length is found to be $0.063\pm 0.002$\,mm for the drying patterns, which translates to about 35 colloid diameters. 
Physically, this is how long the fingers tend to stay aligned, and is therefore a characteristic length between the branching points.
Likewise, we may also find the correlation between angles in the telegraph signal, where the arc-length is the natural variable to compare the decorrelation.
We find that these decorrelation lengths give similar results as the mean arc-lengths, shown in Figure\,\ref{fig:Mean CAL}.

\section{Estimating the colloidal concentration from arc lengths}
\begin{figure}
\centering
\includegraphics[width=0.75\linewidth]{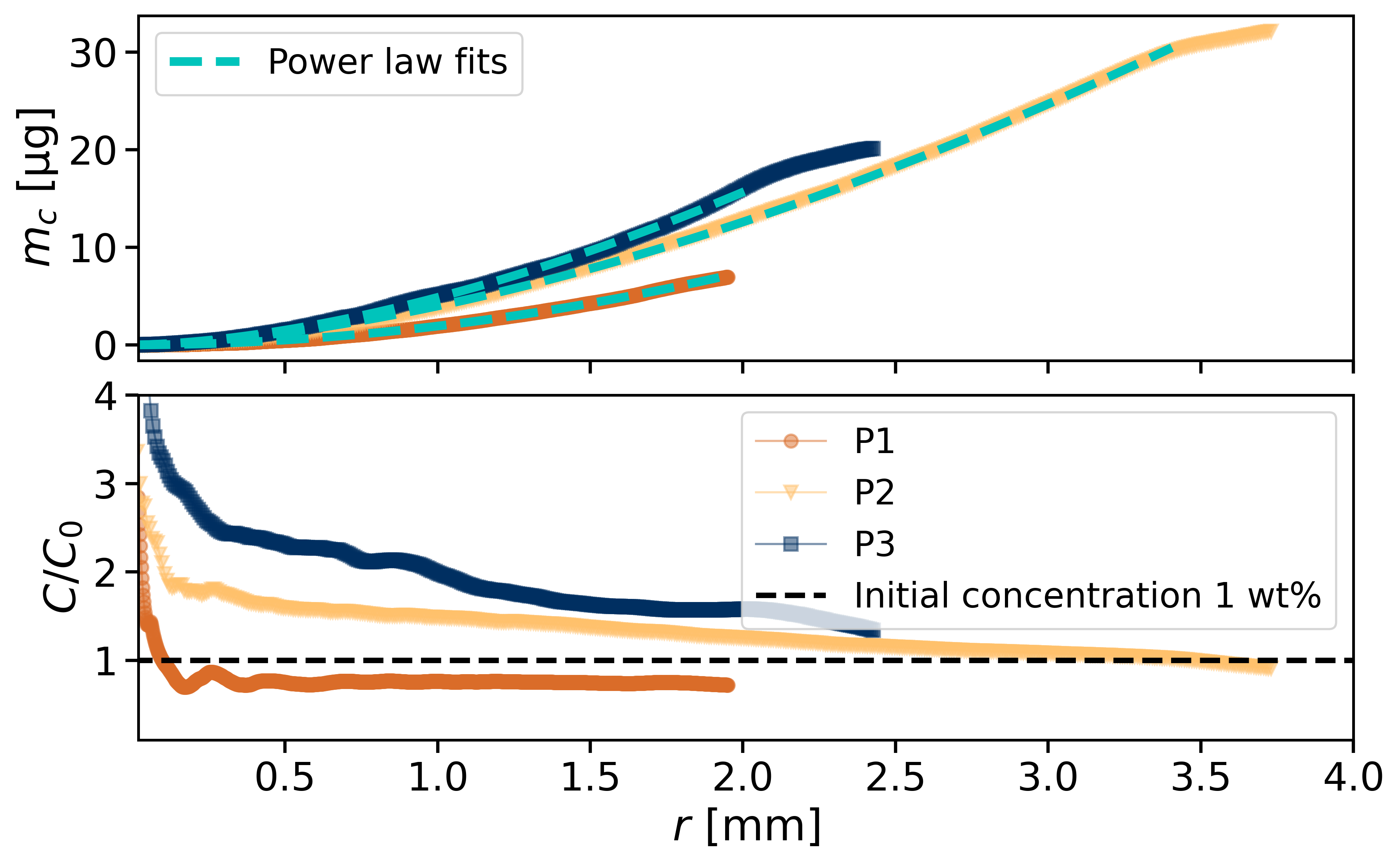}
\caption{A plot of the cumulative colloid mass $m_c(r)$ deposited in the pattern forming stage of the evaporation process. This is found by assuming a uniform colloidal packing fraction throughout the pattern, left in exactly 2 monolayers. We find that the mass deposition scales with a power-law, with exponents found in Table \ref{tbl:fractal_dimensions}. We then relate the deposited mass to the colloid concentration of the droplet as it is evaporating $C(r)$, compared to the initial concentration $C_0=1$\,wt\% for the patterns P1-P3. The initial concentration of the droplet is illustrated with  a dashed black line.}\label{fig:Colloid_concentration}
\end{figure}

When inspecting the fingers in a microscope, it is clear that they are fairly uniform in colloid density across the pattern, while the space between the fingers contain no colloids at all.
The fingers consist of colloids on both the bottom and the top substrates, with no colloids in between.
The bottom layer is packed with slightly more than a monolayer ($1+\varepsilon$ layers), while the packing in the top substrate has a few holes, roughly equalling the excess ({\it i.e.} $1-\varepsilon$ layers).
We may therefore approximate the fingering pattern as exactly 2 monolayers of colloids consisting of randomly packed monodisperse spheres with a density of $\rho_{TPM}=$1.314\,g\,cm$^{-3}$~\cite{van2017preparation}. 
We assume a packing fraction of a monolayer of random close-packed disks $\phi_{2D}=0.772$~\cite{hinrichsen_random_1990}, leading to an effective 3D packing fraction $\phi=\phi_{2D} \cdot V_{sphere}/V_{cylinder} = 0.513$.
The thickness of the monolayer is the colloid diameter $d_c=1.8$\,$\mu$m.
Now it is possible to relate the information of solid fractions $\varphi(r)$ in Figure\,\ref{fig:area fractions} to the deposited colloidal mass at radius $r$ 
\begin{equation}\label{eq:Colloidal_Mass}
    m_c(r) = \rho_{TPM}\cdot 2\phi \cdot d_{c} \int_0^r \varphi(r')\cdot 2\pi r' \mathrm{d}r'.
\end{equation}
This relation between the mass and radius of a pattern is useful for probing the fractality of the patterns. 
A fractal will have a power-law relation between the mass deposition, scaling with its fractal dimension $m_d(r)\sim r^{D_f}$~\cite{Hansen_etal_2022}. 
In comparison, the scaling of the RBP is simply its Euclidean dimension $m_{d, RBP} = \varphi \pi r^2$.
We find that the mass scaling of patterns P1-P3 are well described by power-law fits, meaning the patterns may be described as fractals, with a fractal dimension similar to that of the DLA pattern (see Figure\,\ref{fig:Colloid_concentration}). 
We also measure the fractal dimension with a box-counting method, all summarized in Table \ref{tbl:fractal_dimensions}. 
These values are fairly similar to the fractal dimension of a DLA pattern~\cite{witten_diffusion-limited_1981}, except for the mass fractal dimension of P1, which is closer to the one of a RBP.
\begin{table}
  \caption{Comparison of the fractal dimensions of the patterns P1-P3, measured with two different methods. The mass scaling method is based on a power-law-fit to the mass deposition data, while the box-counting method measures how many boxes $N$ of size $\delta$ we need to cover the full pattern, with the fractal dimension found as $D_f=\ln{N}/\ln{\delta}$.}
  \label{tbl:fractal_dimensions}
  \begin{tabular}{lll}
    \hline
      & Mass scaling dimension & Box-counting dimension  \\
    \hline
    P1 & 1.95 & 1.61 \\
    P2 & 1.67 & 1.68 \\
    P3 & 1.69 & 1.72 \\
    \hline
  \end{tabular}
\end{table}
We may now relate the deposited mass to the colloids deposited in the fingers.
As the droplet shrinks, it can be split into its fingers and its ``inscribed circle'' of radius $r$ as in ref.\,\citenum{Beechey-Newman_etal_2025}. 
The mass of the colloids in this evaporating droplet must be the same mass we find in the fingers between the center and $r$, and can be calculated with equation \eqref{eq:Colloidal_Mass}.
We compute the weight of the water as $m_w(r)=\rho_w\pi r^2 h_{cell}$, with the value $h_{cell}=80$\,$\mu$m. 
Finally, we compute the colloidal concentration in the evaporating droplet as it shrinks $C(r)=\frac{m_c(r)}{m_w(r)+m_c(r)}$, which is compared to the initial colloid concentration $C_0$ and shown in Figure\,\ref{fig:Colloid_concentration}. 
Interestingly, we note that the instability leading to the finger formation occurs at different initial (average) colloid concentrations, at $C(r_{max})/C_0=0.70$ for P1, $0.88$ for P2 and $1.29$ for pattern P3. 
We may also estimate the initial mass of the colloids.
The initial concentration is $1\text{ }wt\%$, and the cell dimensions are $r=6$\,mm and $h_{cell}=80$\,$\mu$m. 
Thus, the total mass of the colloids $m_{c, tot}$ is roughly 100\,$\mu$g. 
The evaporating colloidal systems are found to initially deposit colloids in a uniform monolayer (seen as gray color inside the cell in Figure\,\ref{fig:Experimental Setup}c), until there is a transition to the finger instability~\cite{Beechey-Newman_etal_2025}.
We can assume the droplet {\it either} leaves colloids in the initially forming monolayer, {\it or} in the fingering pattern. 
Thus, by taking $r' = r_{max}$ in equation \eqref{eq:Colloidal_Mass}, we find the mass of colloids going into the fingering pattern.
We compare this weight to the initial colloid weight, and find the percentage of colloids going in the fingers to be 7\,\% for pattern P1, 31\,\% for P2 and 19\,\% for pattern P3.

\section{Using the telegraph signal to find the centre}

The choice of the location of the centre of a roughly radially symmetric pattern is to some extent arbitrary. 
The simplest method to find the centre of mass of a pattern is to locate its center of mass. However, for the colloidal drying patterns, the centre of mass is skewed by the transversal thick fingers forming at the rim of the pattern.
These thick fingers tend to form unevenly, hence the centre of mass is clearly different from any visually selected centre.
If the pattern is a fully connected network, like a DLA, it is possible to follow the fingers towards their centre.
Following the idea of Horton and Strahlers works on complexity in river networks~\cite{Horton1945, Strahler1957}, we may skeletonize a pattern and count the numbers of bifurcations at each node in the skeleton network.
However, this method may provide results that are clearly off the centre, as it a) requires a fully contiguous pattern and b) cause arbitrariness where there are multiple equally ordered points.

\begin{figure*}[!ht]
\centering
\includegraphics[width=0.75\linewidth]{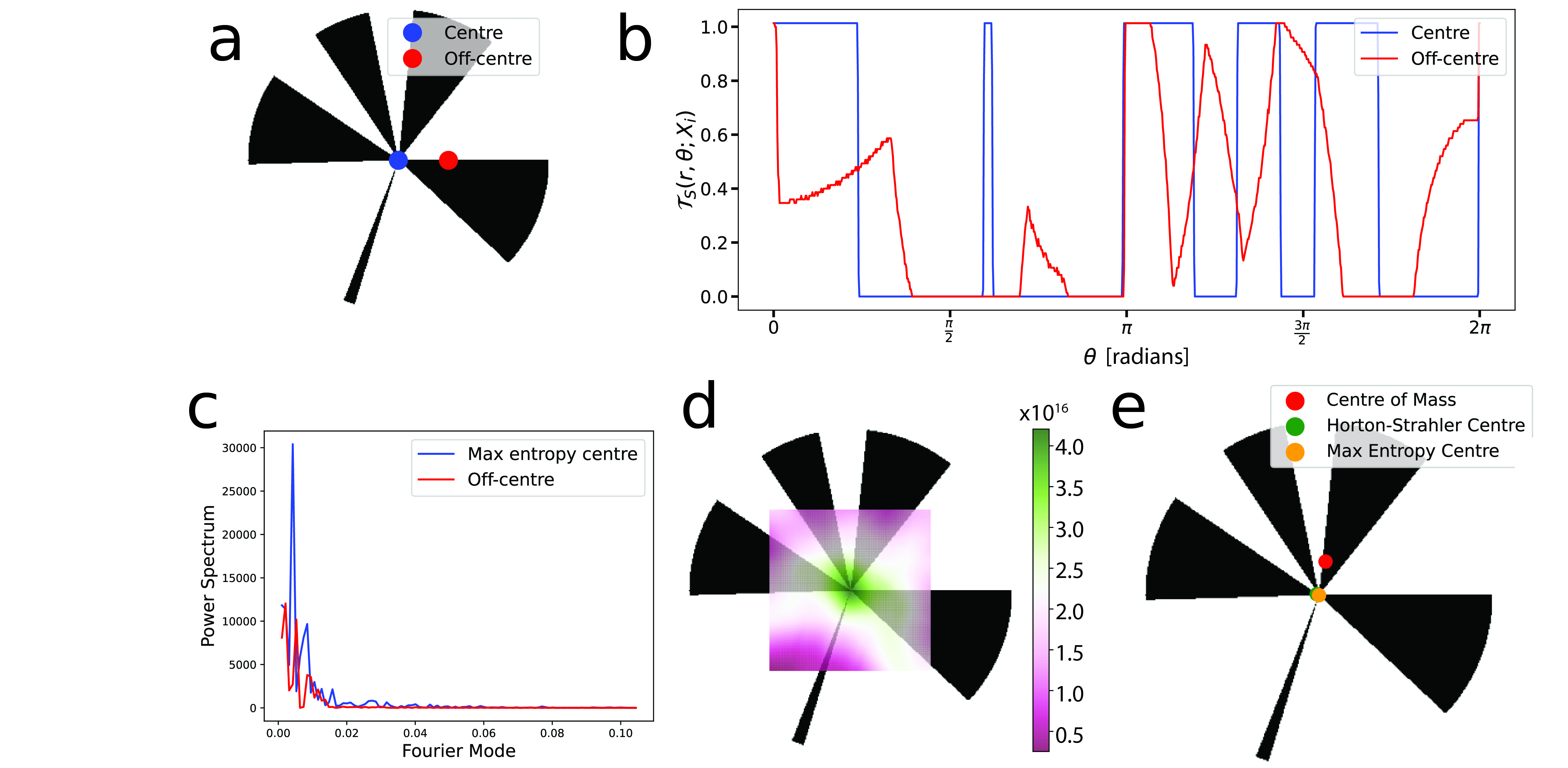}
\caption{(a)-(c): A comparison of two points, where one is in the centre of the Random Birthday Pie pattern and the other is placed 50 pixels to the right, as shown in (a). (b) compares the summed and normalised telegraph signal $\mathcal{T}_S(\theta; X_i)$ from the two points, and (c) is their respective power spectra, which is used to calculate the entropy. (d) shows a map of the calculated entropies at each point in the selected window, and (e) shows the centre calculated with 3 different methods.}
\label{fig:maximal entropy}
\end{figure*}

In order to solve the arbitrariness of deciding the centre, we propose a method using the arc-length distribution. 
First, we define a test centre $X_t$, where we may use any of the mentioned approaches as a first approximation. 
We define a window of $N\times N$ pixels centred around $X_t$. 
We then decide the range of relevant radii, $r\in [r_{\text{min}}, r_{\text{max}}]$. 
For each pixel, $X_i$ within the window, we sum the telegraph signal as
$$\mathcal{T}_S(\theta; X_i) = \frac{1}{r_{\text{max}} - r_{\text{min}}} \int_{r_{\text{min}}}^{r_{\text{max}}} \mathcal{T}(r, \theta; X_i) \mathrm{d}r.$$

We then calculate the power spectrum of the signal as
$G(\hat{\theta})=\int_{0}^{\infty}|\mathcal{T}_S(\theta; X_i)|^2 \mathrm{d}\theta,$
where $\hat{\theta}$ denotes the Fourier mode of the angle.
A clear periodicity in the summed telegraph signal will result in a sharp peak in the power spectrum. 
We therefore use the power spectrum to define an information entropy, as
$S(X_i) = \int_0^\infty G(\hat{\theta}; X_i) \ln{G(\hat{\theta}; X_i)} \mathrm{d}\hat{\theta}.$
The best centre, based on our information entropy, is simply picked as the extremal point $X_c = \max(S(X_i))$.
Figure\,\ref{fig:maximal entropy}a-c shows a comparison of two potential centres for the RBP figure, along with their power spectra. Summing the spectra as an entropy, we obtain the map in Figure\,\ref{fig:maximal entropy}d. In Figure\,\ref{fig:maximal entropy}e the comparison between the maximum entropy, centre of mass and Horton-Strahler approaches shows that our maximum entropy determination of the figure's centre is the most accurate.

The RBP and DLA patterns have well-defined centres, so we may quantify the errors; these are presented in Table\,\ref{tbl:errors_com_entropy}. 

\begin{table}
  \caption{Comparison of the errors we find in the centre of mass and entropy methods. For the DLA we use 20 generated patterns of 100000 particles and a max radius of 1024 pixels. For the Random Birthday Pie we generate 20 random patterns with 10 fingers and a max radius of 150 pixels.}
  \label{tbl:errors_com_entropy}
  \begin{tabular}{lll}
    \hline
      & Centre of Mass & Entropy  \\
    \hline
    DLA   & $9.53$ px & $22.34$ px    \\
    RBP & $16.92$ px & $0.84$ px   \\
    \hline
  \end{tabular}
\end{table}

\begin{figure}
\centering
\includegraphics[width=0.75\linewidth]{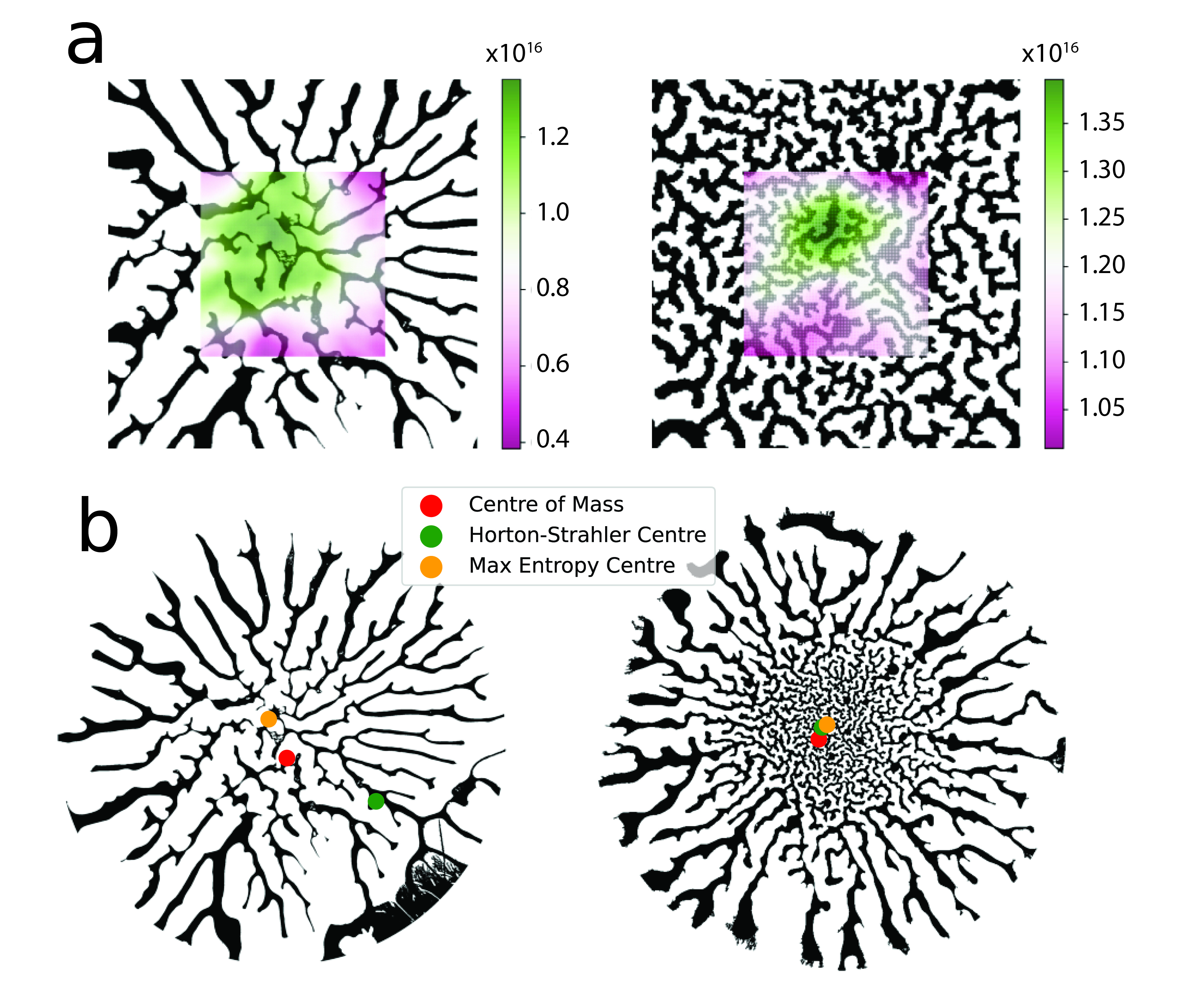}
\caption{(a) Resulting entropy map of patterns P1 and P2, using the method presented in this paper. Green in the scale bar corresponds to the best location of the centre. (b) Resulting centre points, compared with the two alternative methods.}
\label{fig:entropy_map_P1_P2}
\end{figure}

To find the best centre to apply our CAL analysis of samples P1 and P2 are not as straight forward. Depending on experimental parameters such as possible pinning centres on the support surfaces or  on how the samples are levelled horizontally, the apparent centre may not be located at the centre of the sample chamber. However, we can choose a centre by eye and then obtain the maps of the local entropies. These are shown, along with a comparison with other methods in Figure\,\ref{fig:entropy_map_P1_P2}. Clearly, the green region is most representative of the true centre of the labyrinthian pattern of our samples. Hence our maximum entropy approach helps to find the best centre, around which the resulting CAL analysis delivers most reliable results for comparisons with other growth laws.

\section{Conclusions}
We find that extending the traditional chord length distributions to roughly radially symmetric patterns enable us to characterize these patterns in a way that is not feasible with the traditional method.
The distributions may serve as a fingerprint of the patterns, which vary from experiment to experiment, due to the subtle differences in the exact initial conditions.

We find that the colloidal drying patterns P1-P3 fall somewhere between the characteristic DLA and Random Birthday Pie patterns. 
Most importantly, we find the competition between the generation and merging of fingers sustains a fairly stable number of fingers, up to a point where the fingers merge more rapidly. 
We compute the correlation between the telegraph signals, and relate them to a characteristic exponential decay with a persistence length of $\xi=0.063$\,mm.
With a few assumptions, we are also able to relate the information from the arc-lengths to describe how the pattern deposits its colloidal mass and to the colloid concentration of the droplet, as it evaporates. This is also used to characterize the fractality of the patterns, showing the fractal dimension is similar to the one of DLA.
Finally, we use the arc lengths to map an information entropy on the pattern, which we may use to pick a centre where this is not trivial.

Hence, the method can be applied to patterns that are not simply radially invariant or self-similar, but fall somewhere between the two. 

\section{Data Availability}
All the images, methods and Python codes used in this paper is available in a public Zenodo repository~\cite{hennig_2026_18954724}. This includes the method used to generate the synthetic patterns (Diffusion-Limited Aggregation and Random Birthday Pies), the method we used to binarize our experimental images and the full, annotated Python codes we developed to perform our analyses.

\begin{acknowledgement}
This work was supported by the Research Council of Norway through its Center of Excellence funding scheme, project number RCN 262644. The authors would like to thank Daan Frenkel and Benjamin Rotenberg for fruitful discussions and critically reading this manuscript.
\end{acknowledgement}

\bibliography{bibliography}

@article{Mering_Tchoubar_1968,
  title = {{Interpr{\'e}tation de la diffusion centrale des rayons X par les syst{\`e}mes poreux. I}},
  author = {M{\'e}ring, J. and Tchoubar, D.},
  year = {1968},
  month = sep,
  journal = {Journal of Applied Crystallography},
  volume = {1},
  number = {3},
  pages = {153--165},
  publisher = {International Union of Crystallography},
  issn = {0021-8898},
  doi = {10.1107/S0021889868005212},
  urldate = {2025-02-26},
  langid = {french}
}

@article{Zupkauskas_etal_2017,
  title = {Optically Transparent Dense Colloidal Gels},
  author = {Zupkauskas, M and Lan, Y and Joshi, D and Ruff, Z and Eiser, E},
  year = {2017},
  journal = {Chemical Science},
  volume = {8},
  number = {8},
  pages = {5559--5566},
  publisher = {Royal Society of Chemistry}
}

@article{Deegan_etal_1997a,
  title = {Capillary Flow as the Cause of Ring Stains from Dried Liquid Drops},
  author = {Deegan, Robert D and Bakajin, Olgica and Dupont, Todd F and Huber, Greb and Nagel, Sidney R and Witten, Thomas A},
  year = {1997},
  journal = {Nature},
  volume = {389},
  number = {6653},
  pages = {827--829},
  publisher = {Nature Publishing Group UK London}
}

@article{Hu_Larson_2006a,
  title = {Marangoni Effect Reverses Coffee-Ring Depositions},
  author = {Hu, H. and Larson, R. G.},
  year = {2006},
  journal = {The Journal of Physical Chemistry B},
  volume = {110},
  number = {14},
  pages = {7090--7094},
  doi = {10.1021/jp0609232}
}

@article{Abdulsahib2022,
  author = {Abdulsahib, A. A. and Mahmoud, M. A. and Aris, H. and Gunasekaran, S. S. and Mohammed, M. A.},
  year = {2022},
  title = {An Automated Image Segmentation and Useful Feature Extraction Algorithm for Retinal Blood Vessels in Fundus Images},
  journal = {Electronics},
  volume = {11},
  pages = {1295},
  note = {doi:10.3390/electronics11091295.},
}

@article{Majumder_etal_2012,
  title = {Overcoming the ``{{Coffee-Stain}}'' {{Effect}} by {{Compositional Marangoni-Flow-Assisted Drop-Drying}}},
  author = {Majumder, Mainak and Rendall, Clint S. and Eukel, J. Alexander and Wang, James Y. L. and Behabtu, Natnael and Pint, Cary L. and Liu, Tzu-Yu and Orbaek, Alvin W. and Mirri, Francesca and Nam, Jaewook and Barron, Andrew R. and Hauge, Robert H. and Schmidt, Howard K. and Pasquali, Matteo},
  year = {2012},
  month = jun,
  journal = {The Journal of Physical Chemistry B},
  volume = {116},
  number = {22},
  pages = {6536--6542},
  publisher = {American Chemical Society},
  issn = {1520-6106},
  doi = {10.1021/jp3009628},
  urldate = {2024-11-06}
}

@article{li_et_al_2020,
  title={Evaporating droplets on oil-wetted surfaces: suppression of the coffee-stain effect},
  author={Li, Yaxing and Diddens, Christian and Segers, Tim and Wijshoff, Herman and Versluis, Michel and Lohse, Detlef},
  journal={Proceedings of the National Academy of Sciences},
  volume={117},
  number={29},
  pages={16756--16763},
  year={2020},
  publisher={National Academy of Sciences}
}

@article{Beechey-Newman_etal_2025,
author = {Ilaria Beechey-Newman  and Natalya Kizilova  and Andreas Andersen Hennig  and Eirik Grude Flekkøy  and Erika Eiser },
title = {Confined colloidal droplets dry to form circular mazes},
journal = {Proceedings of the National Academy of Sciences},
volume = {122},
number = {32},
pages = {e2508363122},
year = {2025},
doi = {10.1073/pnas.2508363122},
URL = {https://www.pnas.org/doi/abs/10.1073/pnas.2508363122},
eprint = {https://www.pnas.org/doi/pdf/10.1073/pnas.2508363122}}

@incollection{Hansen_etal_2022,
  title = {Physics of {{Flow}} in {{Porous Media}}},
  booktitle = {Physics of {{Flow}} in {{Porous Media}}},
  author = {Hansen, Alex and Flekk{\o}y, Eirik Grude and Feder, Jens},
  year = {2022},
  publisher = {Cambridge University Press},
  address = {Cambridge},
  doi = {10.1017/9781009100717.018},
  urldate = {2024-06-03},
  isbn = {978-1-108-83911-2}
}

@article{Levitz_Tchoubar_1992,
  title = {Disordered Porous Solids : From Chord Distributions to Small Angle Scattering},
  shorttitle = {Disordered Porous Solids},
  author = {Levitz, P. and Tchoubar, D.},
  year = {1992},
  month = jun,
  journal = {Journal de Physique I},
  volume = {2},
  number = {6},
  pages = {771--790},
  issn = {1155-4304, 1286-4862},
  doi = {10.1051/jp1:1992174},
  urldate = {2024-04-09},
  langid = {english}
}

@article{Horton1945,
  author = {Horton, R. E.},
  year = {1945},
  title = {Erosional Development of Streams and Their Drainage Basins; Hydrophysical Approach to Quantitative Morphology},
  journal = {GSA Bulletin},
  volume = {56},
  number = {3},
  pages = {275-370},
  note = {doi:10.1130/0016-7606(1945)56[275:EDOSAT]2.0.CO;2.},
}

@article{Strahler1957,
  author = {Strahler, A. N.},
  year = {1957},
  title = {Quantitative analysis of watershed geomorphology},
  journal = {Transactions American Geophysical Union},
  volume = {33},
  pages = {913-920},
  note = {doi:10.1029/TR038i006p00913.},
}

@article{Otsu_1979,
  title = {A {{Threshold Selection Method}} from {{Gray-Level Histograms}}},
  author = {Otsu, Nobuyuki},
  year = {1979},
  month = jan,
  journal = {IEEE Transactions on Systems, Man, and Cybernetics},
  volume = {9},
  number = {1},
  pages = {62--66},
  issn = {2168-2909},
  doi = {10.1109/TSMC.1979.4310076},
  urldate = {2024-12-16}
}

@article{van2017preparation,
  title={Preparation of colloidal organosilica spheres through spontaneous emulsification},
  author={Van Der Wel, Casper and Bhan, Rohit K and Verweij, Ruben W and Frijters, Hans C and Gong, Zhe and Hollingsworth, Andrew D and Sacanna, Stefano and Kraft, Daniela J},
  journal={Langmuir},
  volume={33},
  number={33},
  pages={8174--8180},
  year={2017},
  publisher={ACS Publications}
}

@article{witten_diffusion-limited_1981,
    title = {Diffusion-{Limited} {Aggregation}, a {Kinetic} {Critical} {Phenomenon}},
    volume = {47},
    url = {https://link.aps.org/doi/10.1103/PhysRevLett.47.1400},
    doi = {10.1103/PhysRevLett.47.1400},
    number = {19},
    urldate = {2026-03-05},
    journal = {Physical Review Letters},
    author = {Witten, T. A. and Sander, L. M.},
    month = nov,
    year = {1981},
    note = {Publisher: American Physical Society},
    pages = {1400--1403},
}

@article{kaufman_parallel_1995,
    title = {Parallel diffusion-limited aggregation},
    volume = {52},
    url = {https://link.aps.org/doi/10.1103/PhysRevE.52.5602},
    doi = {10.1103/PhysRevE.52.5602},
    number = {5},
    urldate = {2026-03-05},
    journal = {Physical Review E},
    author = {Kaufman, Henry and Vespignani, Alessandro and Mandelbrot, Benoit B. and Woog, Lionel},
    month = nov,
    year = {1995},
    note = {Publisher: American Physical Society},
    pages = {5602--5609},
}

@article{hinrichsen_random_1990,
    title = {Random packing of disks in two dimensions},
    volume = {41},
    url = {https://link.aps.org/doi/10.1103/PhysRevA.41.4199},
    doi = {10.1103/PhysRevA.41.4199},
    number = {8},
    urldate = {2026-03-11},
    journal = {Physical Review A},
    author = {Hinrichsen, Einar L. and Feder, Jens and Jøssang, Torstein},
    month = apr,
    year = {1990},
    note = {Publisher: American Physical Society},
    pages = {4199--4209},
}

@dataset{hennig_2026_18954724,
  author       = {Hennig, Andreas and
                  Beechey-Newman, Ilaria and
                  Kizilova, Natalya and
                  Eiser, Erika},
  title        = {Arc-length characterization of finite, radial
                   growth patterns
                  },
  month        = mar,
  year         = 2026,
  publisher    = {Zenodo},
  doi          = {10.5281/zenodo.18954724},
  url          = {https://doi.org/10.5281/zenodo.18954724},
}

\end{document}